\documentclass[aps,prl,twocolumn,superscriptaddress,floatfix,showpacs]{revtex4}
\usepackage{amsmath,euscript,theorem,graphicx,rotating}
\usepackage[letterpaper,hmargin=25mm,vmargin=25mm]{geometry}
\usepackage[latin1]{inputenc}
\usepackage[T1]{fontenc}
\usepackage{txfonts}
\usepackage{url}
\usepackage{array}
\usepackage{xr} % for eXternal References
\usepackage{color}
\usepackage{dcolumn}% Align table columns on decimal point
\usepackage{bm}% bold math
\usepackage{longtable}% To allow splitting tables across multiple pages
\usepackage{xspace}% To preserve spaces after macros
\usepackage{booktabs}

\usepackage{array}
\newcommand{\dr}{{\mathrm d}^{3}{\mathbf r}}
\newcommand{\drp}{{\mathrm d}^{3}{\mathbf r'}}
\newcommand{\rr}{{\mathbf r}}
\newcommand{\RR}{{\mathbf R}}
\newcommand{\rp}{{\mathbf r'}}

\newcommand{\rff}[1]{{\ (\ref{#1})}}
\newcommand{\etal}{\emph{et al.}\xspace}

\newcommand{\fdds}{\ensuremath{\alpha(\rr,\rp|\omega)}\xspace}

\newcommand{\Q}[2]{\ensuremath{\hat{Q}^{#1}_{#2}}\xspace}
\newcommand{\T}[2]{\ensuremath{T^{#1}_{#2}}\xspace}
\newcommand{\A}[2]{\ensuremath{\alpha^{#1}_{#2}}\xspace}

\newcommand{\xhat}{\hat{\mathbf{x}}}
\newcommand{\zhat}{\hat{\mathbf{z}}}

\newcommand{\Ri}{{\rm i}}

\newcommand{\Edisp}[0]{\ensuremath{E^{(2)}_{\rm disp}}\xspace}

\newcommand{\distortion}{\ensuremath{\eta}\xspace}

\newcommand{\HH}[1]{(H$_2$)$_{#1}$\xspace}
\newcommand{\bra}{\ensuremath{\langle}}
\newcommand{\ket}{\ensuremath{\rangle}}
\newcommand{\xx}{{\mathbf x}}

\newcommand{\adag}{{a^\dagger}}

\mathchardef\lt="313C \mathchardef\gt="313E
\mathcode`<="4268 \mathcode`>="5269
\newcommand{\JCP}[0]{J. Chem. Phys.\ }
\newcommand{\JPCA}[0]{J. Phys. Chem. A\ }

\newcommand{\JCTC}[0]{J. Chem. Theory Comput.\ }
\newcommand{\IJQC}[0]{Int. J. Quantum Chem.\ }
\newcommand{\CPL}[0]{Chem. Phys. Lett.\ }

\newcommand{\PR}[0]{Phys. Rev.\ }
\newcommand{\PRA}[0]{Phys. Rev. A\ }

\newcommand{\PRL}[0]{Phys. Rev. Lett.\ }

\newcommand{\MolP}[0]{Mol. Phys.\ }

\newcommand{\PCCP}[0]{Phys. Chem. Chem. Phys.\ }

\begin{document}

\title{Dispersion interactions between semiconducting wires}

\author{Alston J. Misquitta}
\affiliation{Cavendish Laboratory, 19, J J Thomson Avenue
Cambridge CB3 0HE, U.K.}

\author{James Spencer}
\affiliation{University Chemical Laboratory, Lensfield Road, 
Cambridge CB2 1EW, U.K.}
\affiliation{Department of Physics and Thomas Young Centre, 
Imperial College London, Exhibition Road, London SW7 2AZ, U.K}

\author{Anthony J. Stone}
\affiliation{University Chemical Laboratory, Lensfield Road, 
Cambridge CB2 1EW, U.K.}

\author{Ali Alavi}
\affiliation{University Chemical Laboratory, Lensfield Road, 
Cambridge CB2 1EW, U.K.}

\date{\today}

\begin{abstract}
The dispersion energy between extended molecular chains (or equivalently infinite wires) 
with non-zero band gaps is generally assumed to be expressible as a pair-wise sum of 
atom-atom terms which decay as $R^{-6}$. Using a model system of two
parallel wires with a variable band gap, we show that this is not the case. 
The dispersion interaction scales as $z^{-5}$ for large interwire
separations $z$, as expected for an insulator, but as the band gap
decreases the interaction is greatly 
enhanced; while at shorter (but non-overlapping) separations
it approaches a power-law scaling given by $z^{-2}$, \emph{i.e.} the dispersion
interaction expected between \emph{metallic} wires.
We demonstrate that these effects
can be understood from
the increasing length scale of the plasmon modes (charge
fluctuations), and their increasing contribution to the molecular
dipole polarizability and the dispersion interaction,
as the band gaps are reduced. This result 
calls into question methods which invoke locality assumptions in deriving dispersion interactions
between extended small-gap systems.
\end{abstract}

%PACS
%31.10.+z       Theory of electronic structure, electronic transitions, and
%chemical binding
%31.15.-p        Calculations and mathematical techniques in atomic and
%molecular physics
%31.15.ap       Polarizabilities and other atomic and molecular properties
%31.15.bu       Semi-empirical and empirical calculations (differential
%overlap, Hückel, PPP methods, etc.)
%34.20.Gj       Intermolecular and atom-molecule potentials and forces
%73.22.-f       Electronic structure of nanoscale materials and related
%systems
%81.05.Fb       Organic semiconductors


\maketitle

\section{Introduction}
\label{sec:introduction}

Conventionally, the dispersion interaction between two systems is formulated
in terms of correlated fluctuations described by local (frequency-dependent)
polarizabilities which gives rise to the familiar atom--atom 
$-C_6^{ab} R_{ab}^{-6}$ interaction at leading order \cite{Stone:book:96}. 
This form of the interaction has been used with a good measure of success in
studies of many systems, from gases to solids, including complexes
of biological molecules and organic molecules.
These are typically insulators; that is, they have large HOMO--LUMO gaps.

It has been known for a very long time that the {\em additive} atom--atom
form of the dispersion does not hold very well for metallic systems. The classic
case of an atom interacting with a thin metallic surface shows deviations 
from the additive law: the correct treatment of the metal results in a
$R^{-3}$ power law \cite{CasimirP48}, but summing over atom--atom terms
leads to a $R^{-4}$ interaction.
More recently, strong deviations from additivity have been demonstrated in 
the interactions of extended systems of metals or zero band-gap materials 
with at least one nano-scale dimension
\cite{Dobson07a,DobsonWR06,DobsonEtAl01,DobsonEtAl04,DrummondN07}.

The semi-classical picture is useful for understanding the source of the
differences in the metallic and insulating cases. In an insulator, electronic
perturbations decay exponentially with distance. We can therefore treat
electron correlations as being local. At lowest order, these local
fluctuations give rise to instantaneous {\em local} dipoles, 
and the correlation between these local dipoles gives rise to the 
atom--atom $R^{-6}$ interaction.
However, in a zero band-gap material, electronic fluctuations are long-ranged,
particularly if one or two of the dimensions are nano-scale
\cite{Dobson07a}.
It is these long-ranged fluctuations that give rise to the non-additivity
of the dispersion and deviations from the atom--atom $R^{-6}$ interaction.

While the insulating and metallic cases are now well understood, 
little is known of the intermediate, semi-conducting case. 
The nature of the dispersion interaction between extended molecules with
finite but small HOMO--LUMO gaps less than about 0.2 a.u. (5 eV) remains an open
question.
A large number of important nano-molecules fall into this category, such as
carbon nano-tubes and the `lander'-type molecules that are used as organic 
conductors. The electronic structure of materials made of these molecules
depends strongly on structures they assume in the bulk, often via
self-assembly. Consequently it is very important to understand exactly how
these molecules interact.
The dispersion interaction is the dominant source of attraction between these
$\pi$-conjugated systems. 
For want of a clearer understanding, many studies assume the usual insulating
case for the atom--atom $R^{-6}$ form of this interaction. 
As we shall show in this article, this is both qualitatively and 
quantitatively incorrect for semiconducting molecules. 

A word of clarification: we use the term `non-additive' here to
describe deviations from the additive atom-atom picture of the
dispersion interaction between two molecules. It is also commonly used
to refer to the deviation from pair additivity seen in the
interactions of three or more distinct molecules, but we are not
concerned with such effects here, except for a brief comment in the
Discussion.

\section{Interacting wires using H{\"u}ckel (Tight-binding) theory}
\label{sec:hueckel}

The dispersion energy appears at second-order in intermolecular perturbation
theory and is formally expressed in terms of the exact eigenstates and
eigenenergies of the non-interacting systems \cite{Stone:book:96}.
In a mean-field theory the dispersion energy between two subsystems ($A$ and $B$), 
can be expressed as a sum over the single-electron wavefunctions localised 
to each subsystem:
\begin{equation}
\Edisp = \sum_{i\in A, j\in B} \sum_{a\in A, b\in B} 
      \frac{|\bra ij |r_{12}^{-1}| ab \ket|^2}
           {\varepsilon_i + \varepsilon_j - \varepsilon_a - \varepsilon_b},  \label{edisp1}
\end{equation}
where $i,j$ ($a,b$) are occupied (virtual) single-particle wavefunctions 
in either subsystem $A$ or $B$,
and $\varepsilon_i$ is the eigenvalue of the $i$-th wavefunction. Assuming two parallel 
wires, aligned parallel to the $x$ axis and separated by a distance
$z$, the integral which appears above has the form: 
\begin{equation}
\bra ij|r_{12}^{-1}|ab\ket = \int 
\frac{
\psi^*_i(x_1)\psi^*_j(x_2)\psi_a(x_1)\psi_b(x_2)}{|x_{12}\xhat+z\zhat|}dx_1dx_2 
\end{equation}
and is the Coulomb interaction between the charge density $\psi_i^*\psi_a$ due to excitation $i\rightarrow a$ 
in subsystem $A$ with $\psi_j^* \psi_b$ in $B$.

In periodic systems, the wavefunction can be expressed in Bloch form, i.e.\ $\psi_j(x)=e^{ik_jx}u_j(x)$, and so 
the co-density $\psi_i^*(x)\psi_a(x)=e^{i(k_a-k_i)x}u_i^*(x)u_a(x)$
can have a long-wavelength modulation, whose electrostatic field will
not be well described 
by a multipole expansion at separations comparable to this length scale. Furthermore, in small-gap systems, 
these excitations have plasmon character and contribute significantly to 
\Edisp, as we discuss below.

H{\"u}ckel (tight-binding) theory gives us a convenient formalism for
evaluating the above expression for interactions between two one-dimensional wires. We considered a two-band model Hamiltonian of the form 
\begin{equation}
H=\sum_{i}^n (\beta \adag_{2i} a_{2i-1} + \beta^\prime \adag_{2i+1} a_{2i} + h.c.),
\end{equation}
where $\beta,\beta^\prime$ are alternating bond-strengths between adjacent sites, as would be encountered in  
a chain of \HH{n} or a $\pi$-conjugated polyene. 
We computed the interactions between two such parallel wires, each  consisting of $2n$ identical atoms, equally spaced at intervals of $d$, such that the unit cell is of length $2d$ and contains two atoms.
Assuming periodic boundary conditions over a crystal cell of length $2dn$, then there are two bands per wavevector.  The single-particle wavefunctions and energies of sych a system can be found analytically.  The band structure is given by
\begin{equation}
\epsilon(k) = \pm\left|\, \beta e^{ikd} + \beta^\prime e^{-ikd}\right|
\end{equation}
and the set of wavevectors by
\begin{equation}
k = \frac{\pi j}{nd}; \quad j = -\frac{n}{2} + 1, -\frac{n}{2} + 2, \cdots , \frac{n}{2}.
\end{equation}
$\beta=\beta^\prime$ corresponds to a uniform wire with energy eigenvalues given 
by $\epsilon(k)=\pm 2\beta |\cos(kd)|$, which in the limit of large $n$ has a vanishing band-gap at half-filling (i.e.\ is a metal). The opposite limit ($\beta^\prime=0$) corresponds to a chain of isolated 
dimers with energy eigenvalues $\epsilon(k)=\pm\beta$, independent of wavevector, and 
corresponds to the perfect insulator. By varying the ratio $\beta^\prime/\beta$ between
0 and 1, we can very conveniently probe the dispersion interaction in the intermediate (semiconducting) 
regime, with the band-gap given by $\Delta E_g=2(\beta-\beta^\prime)$.   
The required integrals can be evaluated \cite{JSpencer:thesis:09} as:
\begin{equation}
\bra ij|r_{12}^{-1}|ab\ket = \frac{1}{nd} \sum_G K_0(G z) Y_{ia}(G) Y_{jb}(k_i+k_j-k_a-k_b-G)
\end{equation}
where $\{G\}$ are reciprocal lattice vectors of the primitive unit cell and
$Y_{ia}(G)$ is the (analytic) Fourier transform of $u_i^*u_a$ obtained by
assuming highly-localised basis functions on each atom.  $K_0$ is the zero-th
order modified Bessel function of the second kind and is the Fourier transform
of the potential generated by a one-dimensional lattice at a field point at $z$
from the lattice\cite{Epstein03,Epstein07, Tosi}.  $K_0$ decays exponentially
with increasingly large arguments and so typically only the 10 smallest
reciprocal lattice vectors need to be included in the summation in order to
obtain converged results.

The calculations presented here use 8401 Monkhorst-Pack $k$-points to sample the Brillouin
zone, which corresponds in real-space to wires consisting of 16802 sites per crystal cell.  Such fine k-point sampling was
required to obtained converged results with respect to system size.
\begin{figure}
\includegraphics[width=0.5\textwidth,clip]{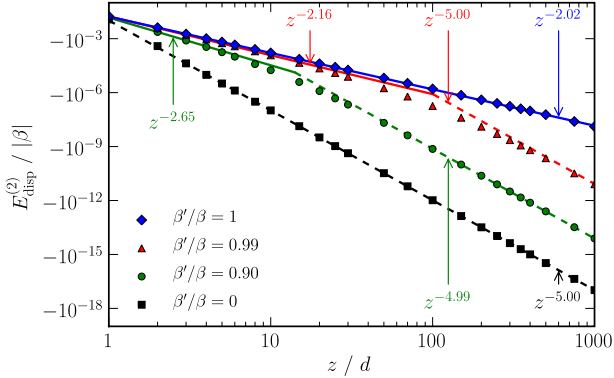}
\caption[Peierls' wire results]{The second-order dispersion energy of parallel
wires modelled using H\"{u}ckel theory obtained from eq.\rff{edisp1}.  The bonding interaction is of
alternating strength and is controlled by the ratio $\beta'/\beta$. The lines are numerical power-law fits to the 
small-$z$ and large-$z$ data points.  In the metallic ($\beta=\beta^\prime$) and perfect insulating ($\beta^\prime=0$) cases, a single power-law 
fits the whole range of $z$, with the expected exponents $-2$ and $-5$ respectively. 
The intermediate cases show a cross-over between these two limiting regimes as $z$ is varied.
\label{fig:peierls_results} 
}
\end{figure}

As shown in fig.\ \ref{fig:peierls_results}, the dispersion interaction between
two uniform wires ($\beta^\prime/\beta=1$) varies as $\sim z^{-2}$ across the whole range of $z$, 
in good agreement with previously obtained results via 
RPA\cite{Dobson07a} and QMC\cite{DrummondN07} for metallic  
wires.  The interaction 
between chains of isolated dimers ($\beta^\prime=0$) varies as $z^{-5}$, as would be expected 
from the conventional picture of dispersion interactions. In the semiconducting regime, 
there is no unique exponent which characterises the dispersion interaction over all $z$. However, at small separations $z$, 
it tends to the metallic behaviour ($z^{-2}$), whereas large $z$ it tends to insulating behaviour $z^{-5}$.    

\section{Interacting H$_2$ chains using SAPT(DFT)}
\label{sec:h2_chains}

The principal drawback of the H{\"u}ckel model is the lack
of electron correlation, and hence screening, within each
subsystem.  For more realistic calculations we can use the
symmetry-adapted perturbation theory based on density-functional theory
\cite{MisquittaJS03,HesselmannJS05,MisquittaPJS05b} (SAPT(DFT)), where the
dispersion energy is evaluated not via a sum-over-states but through a
coupled Kohn--Sham formulation based on the density response functions
of the interacting molecules\cite{LonguetHiggins65}:
\begin{align}
\Edisp &= -\frac{1}{2\pi}\int_0^\infty dw
  \int d\rr_1 d\rr_1' d\rr_2 d\rr_2' \notag\\
    &\qquad \frac{\alpha_A(\rr_1,\rr'_1;\Ri w)\alpha_B(\rr_2,\rr'_2;\Ri w)}
           {|\rr_1-\rr_2||\rr_1'-\rr_2'|} 
           \label{eq:e2disp}
\end{align}
where $\alpha_{A/B}(\rr,\rr';w)$ are the frequency-dependent density response
functions of the molecules which describe the propagation of a 
frequency-dependent perturbation in a molecule, from one point to 
another, within linear-response theory. 
The second-order dispersion energy is exact in this formulation if
exchange effects are neglected. 

\begin{figure}
  \includegraphics[width=0.45\textwidth,clip]{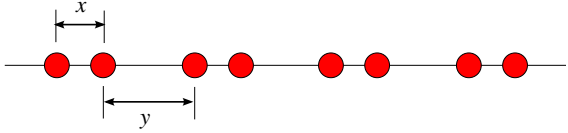}
  \caption[H2n-chains]{
  Distorted \HH{n}-chains. The distortion parameter is defined as $\distortion=y/x$.
  In all our chains we have chosen $x = 1.4487$ a.u.
  \label{fig:H2n-chains}
  }
\end{figure}

We have studied the interactions between two parallel {\it finite} 
\HH{n} (fig.\ \ref{fig:H2n-chains}) chains with $n=16$ and $32$ and
distortion parameters $\distortion = 2.0, 1.5, 1.25$ and $1.0$, where
\distortion is the ratio of the alternate bond lengths. 
The HOMO-LUMO gap of a hydrogen chain can be modified by simply introducing a
difference in alternate bond lengths.
The distortion parameter \distortion gives us a convenient way to control the
electronic structure of the chain and allows us to study the interactions
between insulating, semiconducting and (near) metallic chains in one framework.

The frequency-dependent density response functions that appear in
eq.\rff{eq:e2disp} were evaluated with coupled Kohn--Sham perturbation
theory, using the PBE0 functional with a hybrid
kernel consisting of 75\% adiabatic LDA and 25\% coupled Hartree-Fock.
The accuracy of this approach for the dispersion energy of small molecules
surpasses that of M{\o}ller-Plesset perturbation theory and rivals that of 
coupled-cluster methods \cite{MisquittaPJS05b,HesselmannJ03b,HesselmannJS05,PodeszwaS05}.
Furthermore, there is no qualitative change in our results when the fraction of 
HF exchange is increased, so the results presented here 
are unlikely to be artifacts of the shortcomings of coupled Kohn--Sham
perturbation theory \cite{ChampagnePvGBSS-GRK98,SekinoMKH07}.
All calculations used the cc-pVDZ basis which will result in a significant
underestimation of the strength of the contribution to the dispersion energy
that arises from transverse polarization, but should better describe the
contribution arising from the longitudinal polarization. Since it is the 
longitudinal polarization that is most important in these systems,
we expect our results to be qualitatively correct.

\begin{figure}
  \includegraphics[width=0.5\textwidth,clip]{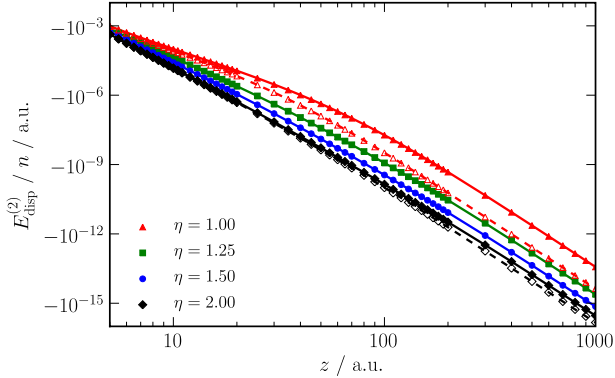}
  \caption[Dispersion H2n-chains]{
  The second-order dispersion energy of pairs of parallel \HH{n} chains.
  Energies have been normalized by the number of H$_2$ units $n$. The solid
  lines are dispersion energies for chains with $n=32$ and the dashed lines
  for chains with $n=16$ (only \distortion=1.0 and 2.0).
  \label{fig:disp-H2n-chains}
  }
\end{figure}

Dispersion energies are displayed in fig.\ \ref{fig:disp-H2n-chains} and
the effective power laws for the physically important separations 
between 6 and 20 a.u.\ are shown in table \ref{tab:power-laws} together with
HOMO--LUMO gaps. 
For insulating chains with an additive dispersion interaction we would expect
two regimes determined by chain length $L$: for $z \ll L$ we should expect
the infinite chain result, i.e., the effective dispersion interaction should
decay as $z^{-5}$, and for $z \gg L$ we should recover the usual 
$z^{-6}$ power law.
This is what we see with the chains with the largest distortion 
parameter, $\distortion=2.0$, which exhibit large HOMO--LUMO gaps. 
As the distortion parameter $\distortion$ approaches unity and the
HOMO--LUMO gap
consequently decreases, the deviation from the additive insulating case
becomes increasingly apparent, and for the longest of the chains considered
here the dispersion interaction is significantly enhanced.  For example at
$40$ a.u. (roughly half the chain length) an $\distortion=1$ chain has
a dispersion interaction two orders of magnitude larger than an 
$\distortion=2$ chain.  Additionally, we see increasing finite-size effects: the
power laws for the $n=16$ and $32$ chains, which were very similar for
$\distortion=2.0$, are considerably different for $\distortion=1.0$.

\begin{table}
\caption[Power laws]{
Power-law behaviour of \Edisp fitted to the form $z^{-x}$ in the region from
6 to 20 a.u. 
Beyond 600 a.u.\ all chains exhibit dispersion energies with the $z^{-6}$
power law because of their finite length. The HOMO--LUMO gaps, $\Delta E_g$, have been calculated from the Kohn--Sham
eigenvalues. Energies are in atomic units.
\label{tab:power-laws}
}
\begin{ruledtabular}
\begin{tabular}{lcccc}
$\distortion$ 
        &\multicolumn{2}{c}{ \HH{16}} 
                           &\multicolumn{2}{c}{ \HH{32}} \\
        &  $\Delta E_g$  &  $x$   &  $\Delta E_g$  &  $x$   \\
   \midrule
2.0     &  0.370         &  4.89  &  0.366         &  4.84  \\
1.5     &  0.280         &  4.58  &  0.270         &  4.50  \\
1.25    &  0.202         &  4.20  &  0.183         &  4.09  \\
1.0     &  0.099         &  3.52  &  0.057         &  3.17  \\
\end{tabular}
\end{ruledtabular}
\end{table}

\section{Multipole expansion}
\label{sec:multipole}

We can understand these effects in terms of the multipole expansion. The
multipole form of the dispersion energy is usually formulated in terms of
(local) atomic polarizabilities.  This leads
to the usual $R^{-6}$ atom--atom interaction (at leading order) and would not
account for the anomalous dispersion power laws that we have observed in the
chains.  However the complete distributed-polarizability description is
non-local: that is, it describes the change in multipole moments at one atom in
response to a change in electrostatic fields at another.

We obtain the multipole form of eq.\rff{eq:e2disp} by expanding the Coulomb
terms using the distributed form of the multipole expansion
$
   |\rr-\rp|^{-1} = \Q{a}{t} \T{ab}{tu} \Q{b}{u}
$
where $a$ and $b$ denote (atomic) sites and $t$ and $u$ are multipole
indices --- 00 for the
charge, 10, $11c$ and $11s$ for the components of the dipole, and so on.
\Q{a}{t} is the multipole moment operator for moment $t$ of site $a$ and
\T{ab}{tu} are the interaction tensors that contain the distance and 
angular dependence (see Ref.\cite{Stone:book:96} for details). 
Inserting this expansion in eq.\rff{eq:e2disp} we obtain the multipole form
for the dispersion energy:
%Using this full non-local description, the dispersion energy becomes\cite[p.140]{Stone:book:96} 
\begin{equation}
\Edisp = -\frac{1}{2\pi}
         \T{ab}{tu} \T{a'b'}{t'u'}
           \int_0^\infty \A{aa'}{tt'}(\Ri w) \A{bb'}{uu'}(\Ri w) dw.
           \label{eq:e2disp-asymp}
\end{equation}
Here $\T{ab}{tu}$ is the interaction function between multipole $t$ on site $a$ in subsystem $A$
and multipole $u$ on site $b$ in $B$, while $\A{aa'}{tt'}$ are the frequency-dependent
non-local polarizabilities for sites $a$ and $a'$, and may be expressed in
terms of the frequency-dependent density susceptibility as\cite{MisquittaS06}
\begin{equation}
  \A{aa'}{tt'}(\omega) = \int_a\int_{a'} \Q{a}{t}(\rr) \fdds \Q{a'}{t'}(\rp) \dr\drp.
  \label{eq:pol}
\end{equation}
%In these equations the multipole indices $t, u$, etc., are 00 for the
%charge, 10, $11c$ or $11s$ for the components of the dipole, and so on. 
There are no assumptions made in deriving eq.\rff{eq:e2disp-asymp},
other than that spheres enclosing the atomic charge densities on different
molecules do not overlap.
%, but this form of the multipole expansion
%involves a quadruple sum over sites.
We have calculated distributed polarizabilities using the constrained
density-fitting algorithm \cite{MisquittaS06}.

Eq.\rff{eq:e2disp-asymp} involves a quadruple sum over sites and is
therefore computationally demanding.
To reduce this cost, we normally make a simplification by {\em localizing} 
the non-local polarizabilities. That is, the non-local
polarizabilities --- the $\A{aa'}{tt'}(\omega)$ with $a\ne a'$ --- 
are transformed onto one or other site using the
multipole expansion \cite{LeSueurS94,LillestolenW07}. 
This transformation is possible only if the non-local
terms decay fast enough with inter-site distance. If not, the multipole
expansion used in the transformation diverges and the localization is no
longer possible. 
As we shall see, this is precisely what happens in the \HH{n} chains as the
distortion parameter $\distortion$ decreases.

\begin{figure*}
\includegraphics[width=0.9\textwidth,viewport=0 0 600 380,clip]{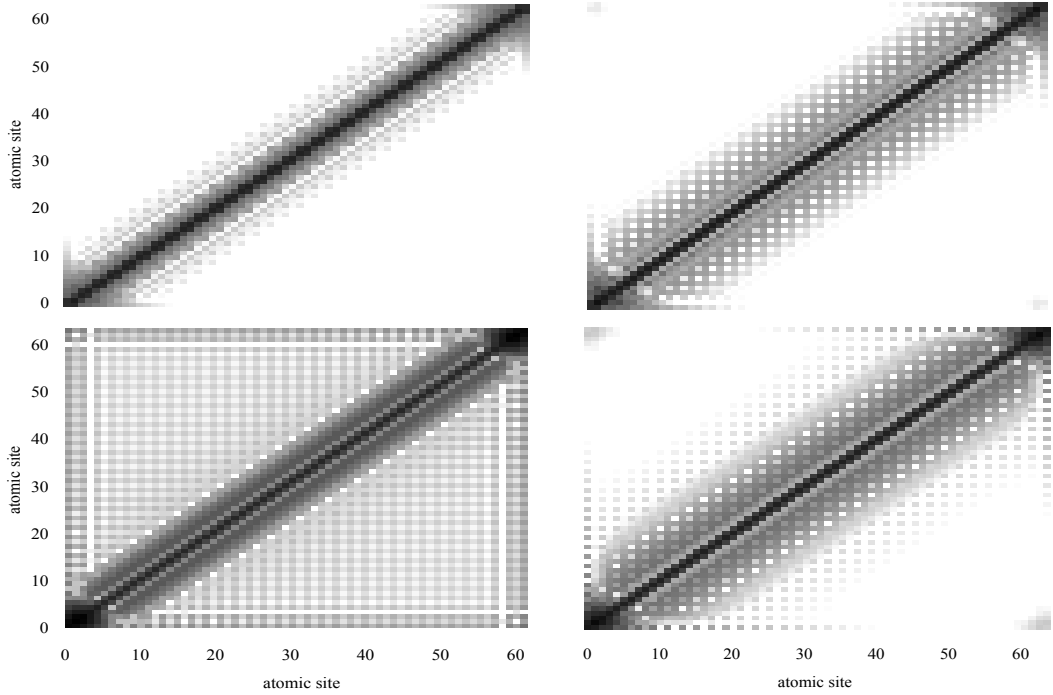}
\caption[Charge-flow polarizabilities]{
Matrix representation of the charge-flow polarizability matrix \A{aa'}{00,00}
for \HH{32} chains with $\distortion=2,1.5,1.25$ and $1$ (clockwise, from top left).
Sites $a$ and $a'$ are represented along the $x$ and $y$ axes.
Because these terms span a number of orders of magnitude of both signs we
have plotted $\ln(|\A{aa'}{00,00}|)$.
Colour scheme: Black rectangles correspond to terms of order 1 a.u.\ and
white rectangles to terms of order $10^{-3}$ a.u.
\label{fig:charge-flow}
}
\end{figure*}

An important feature of the non-local polarizability description
is the presence of {\em charge-flow}
polarizabilities---terms with $t$ or $u=00$ that describe the flow of
charge in a molecule---which are usually small and die off quickly with distance. 
The lowest rank charge-flow polarizability is \A{aa'}{00,00}:
if $V^{a}$ is the potential at site $a$, the change in charge at site 
$a$ is given by $\Delta\Q{a}{00} = - \sum_{a'} \A{aa'}{00,00} (V^{a'}-V^a)$. 
For a system with a large HOMO--LUMO gap the charge-flow terms are expected to be short-range.
The charge-flow polarizabilities of the \HH{32} chain with $\distortion=2.0$ can be
empirically modelled reasonably well with an exponential, that is, 
\begin{equation}
  \A{aa'}{00,00} \sim e^{-\gamma |r_{aa'}|}, 
    \label{eq:chargeflow-decay}
\end{equation}
where $r_{aa'}$ is the intersite distance and $\gamma \approx 0.5$ a.u. 
In this case, the charge-flow polarizabilities drop by
more than an order of magnitude within a few bonds. This is illustrated in
fig.\ \ref{fig:charge-flow}.
However, as we reduce the distortion parameter $\distortion$ the charge-flow
polarizabilities decay more and more slowly with site-site distance, until,
at $\distortion=1.0$, they span the entire length of the chain. 
Even for the $\distortion=1.5$ chain, these non-local charge-flow
polarizabilities can no longer be localized without incurring a significant
error, but for the chains with $\distortion=1.25$ and $1.0$ localization
results in a qualitatively incorrect physical picture.

Moreover the charge-flow polarizabilities contribute to the
dipole--dipole polarizability in the direction of the chain, and this
contribution increases dramatically as the band-gap
decreases. For a 64-atom H$2$ chain with $\distortion=2$, the static dipole
polarizability $\alpha_{xx}=\alpha_{11c,11c}$ parallel to the chain,
calculated as described above, is
about 410 a.u., of which 180 a.u. is contributed by charge-flows. When
$\distortion=1$, so that the H atoms are equally spaced, $\alpha_{xx}$ is
about 11350 a.u., larger by about two 
orders of magnitude, and all of this increase is attributable to the
charge-flow effects.

\begin{figure}
  \includegraphics[width=0.5\textwidth,clip]{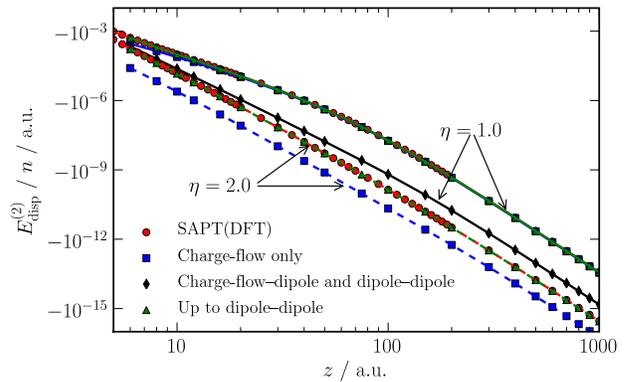}
  \caption[Dispersion from polarizabilities of H2n-chains]{
  The second-order dispersion energy calculated using the 
  polarizabilities of \HH{32} chains with distortion parameters 
  $\distortion=1$ and $2$. 
  The contribution from the charge-flow polarizabilities,
  \A{aa'}{00,00}, is displayed separately from the sum of contributions up to
  the dipole-dipole polarizabilities.
  \label{fig:disp-NLpol-H2n-chains}
  }
\end{figure}

We have calculated the dispersion energy using eq.\rff{eq:e2disp-asymp}
with distributed polarizabilities, including terms to rank~1. Higher ranking
terms can be included, but these are not needed for the hydrogen chains.
These results are presented in fig.\ \ref{fig:disp-NLpol-H2n-chains} for 
the \HH{32} chains with $\distortion=2.0$ and $1.0$. 
First of all, consider the insulating chain ($\distortion=2.0$): the
charge-flow polarizabilities alone severely underestimate the dispersion
energy, but when terms up to rank~1 are included the agreement of 
eq.\rff{eq:e2disp-asymp} with the non-expanded SAPT(DFT) value for \Edisp
is excellent for all inter-chain separations shown in the figure.
However, at $\distortion=1$, the dispersion interaction is entirely dominated   
by the pure charge-flow (i.e. \A{aa'}{00,00}) terms, the higher order terms
involving the dipole polarisabilities making a negligible contribution. 
Since the overall dispersion interaction is greatly enhanced as \distortion is 
reduced, this implies that these pure charge-flow terms enhance the 
dispersion interaction in this limit. 
%If such terms neglected, as it would be in the classic dipole-dipole
%approach, the dispersion interaction would be severely underestimated.  

Why do the charge-flow terms result in the anomalous power laws?
The lowest rank charge-flow terms, \A{aa'}{00,00}, that appear in
eq.\rff{eq:e2disp-asymp} are associated with $T$-functions for the
charge-charge interaction: $\T{ab}{00,00} = R_{ab}^{-1}$, where 
$R_{ab}$ is the distance between site $a$ on one wire and site $b$ on the other.
If the wires are separated by distance vector $\RR=(0,0,z)$ and $\xx_a$ and 
$\xx_b$ are distance vectors for sites $a$ and $b$ along the chains then 
$\RR_{ab} = \RR - (\xx_a-\xx_b)$.
The dispersion energy arising from just the charge-flow terms is
\begin{align}
  \Edisp(00,00) & = - \frac{1}{2\pi} \sum_{aa'}\sum_{bb'} 
                \frac{1}{R_{ab}}\frac{1}{R_{a'b'}}  \nonumber \\
            & \times \int_0^{\infty} \A{aa'}{00,00}(\Ri w) \A{bb'}{00,00}(\Ri w) dw.
            \label{eq:e2disp-chargeflow}
\end{align}
In contrast to the dipole-dipole polarizabilities, the charge-flow
polarizabilities satisfy the sum-rule $\sum_{a'} \A{aa'}{t,00}(w)=0$, 
which is a direct consequence of the charge conservation requirement:
$\int \alpha_A(\rr,\rr'; w) \drp = 0$.
This leads to cancellation between the charge-flow contributions to the
total dispersion energy, and to an $R^{-6}$ distance dependence at
long range, but the cancellation is incomplete at short range, and
terms in $R^{-n}, 2\le n\le 5$ also occur. To see
how this arises, consider the following length scales:
(1) $L_c = 1/\gamma$, which is a measure of the extent of the charge
fluctuations determined by the exponential decay of the charge-flow 
polarizabilities assumed in eq.\rff{eq:chargeflow-decay}, and
(2) $L$, the chain length.
\begin{itemize}
\item
  $z \leq L_c$: Here the charge-flow terms dominate and contribute $R^{-2}$ terms
  to the dispersion energy. From fig.\ \ref{fig:charge-flow} we see that $L_c$
  is largest for the near-metallic wire for which we see the effects of these
  terms (fig.\ \ref{fig:disp-NLpol-H2n-chains}) to $z \sim 60$ a.u.\ 
\item 
  $L_c \ll z \lt L$: In this region the extent of the charge fluctuations is
  small compared with $R_{ab}$ and only those $R_{a'b'}$ close to $R_{ab}$ are
  important. We can therefore expand $R_{a'b'}$ about $R_{ab}$
  using the multipole expansion. The leading term in this expansion is
  \begin{align}
  & \Edisp(00,00) \approx - \frac{1}{2\pi} \sum_{ab}
                \frac{1}{R_{ab}}\frac{1}{R_{ab}}  \nonumber \\
            &\qquad \times \int_0^{\infty} 
            \left( \sum_{a'} \A{aa'}{00,00}(\Ri w)\right)
            \left( \sum_{b'} \A{bb'}{00,00}(\Ri w)\right)
            dw 
            \nonumber \\
            & = 0
  \end{align}
  where we have used the charge-flow sum rule. So we see that the $R^{-2}$
  contributions sum to zero in this region. However, the higher-order contributions 
  are non-zero.
\item 
  $L \ll z$: In this limit both $R_{ab}$ and $R_{a'b'}$ can be expanded in a
  multipole expansion about $\RR=(0,0,z)$: 
  $R_{ab}^{-1} = |\RR - (\xx_a-\xx_b)|^{-1} = |\RR - \xx_{ab}|^{-1} \approx
  z^{-1} - \frac{1}{2} x_{ab}^2 z^{-3}$, and likewise for $R_{a'b'}^{-1}$.
  Once again, the leading terms vanish because of the sum rule, leaving 
  only the effective dipole-dipole contribution:
  \begin{align}
  \Edisp(00,00) & \approx - \frac{1}{2\pi} \frac{1}{4z^6} \sum_{aa'}
  \sum_{bb'} x_{ab}^2 x_{a'b'}^{2}  \nonumber \\
    & \times \int_0^{\infty} \A{aa'}{00,00}(\Ri w) \A{bb'}{00,00}(\Ri w) dw
    \nonumber \\
                & \equiv - \frac{C_6(00,00)}{z^6}.
  \end{align}
  This explains the large-$z$ charge-flow contribution to the dispersion
  energy shown in fig.\ \ref{fig:disp-NLpol-H2n-chains}.
  Notice that for the near-metallic wires, this contribution to the total
  molecular $C_6$ coefficient dominates that from the dipole-dipole
  polarizabilities, but for the large-gap wires the opposite is true.
\end{itemize}
In short, the charge-flow polarizabilities give rise to the
changes in power-law of the dispersion energy and also contribute to an
enhancement of the effective $C_6$ coefficient, applicable at long range.

\section{Discussion}
\label{sec:discussion}

The physical picture which emerges from both the H\"{u}ckel approach and the
\emph{ab initio} calculations is the following.
\begin{itemize}
\item
  In systems with a finite but small gap, spontaneous charge-fluctuations 
  (plasmon modes in the infinite case)
  introduce a secondary length-scale intermediate in size to the interatomic
  distance and the system size. This length scale grows as the gap gets smaller. 
\item
  For separations $z$ small compared with this length scale, 
  the dispersion energy arising from the correlated fluctuations has
  metallic character.
\item
  At $z$ that are large compared with the length scale of the fluctuations,
  the dispersion can be described using London's dipole approximation, 
  giving $z^{-5}$ behaviour, but the magnitude of the fluctuations now depends
  strongly on the band gap, leading to orders of magnitude enhancement over
  the insulator case.
\item 
  In small-gap systems these fluctuations give rise to a strong non-additivity
  in the polarizability. For example, the ratio of the longitudinal static
  polarizabilities for the near-metallic and insulating \HH{32} chains is
  28. The dispersion energy is proportional to the square of the
  polarizability, that is, 784. This is roughly the ratio of the dispersion
  energies for these two cases but only at separations $z$ much greater than
  the chain length. Any attempt to extrapolate this result to shorter
  distances results in a severe overestimation of the dispersion energy.
\item 
  This effect is not a consequence of retardation (we use the non-relativistic
  Hamiltonian) or damping (charge-density overlap is negligible). It
  originates from the complex behaviour of the non-local charge-flow
  polarizabilities. These are terms of rank zero that describe charge
  fluctuations in the system and are associated with a delocalized
  exchange-correlation hole \cite{Angyan07a,Angyan09a}.
  This delocalization can be quantified using the
  localization tensor \cite{RestaS99a,Angyan09a} and may give us a quantitative
  method for defining the charge fluctuation length-scale, $L_c$. We are
  currently investigating this possibility.
\item 
  We have demonstrated that both the change in power-law with
  distance and the enhancement of the dispersion energy can be understood using
  non-local polarizability models containing charge-flow polarizabilities. 
\item 
  It should come as no surprise that these effects are also strongly dependent
  on system size (see fig.\ref{fig:disp-H2n-chains}). 
\end{itemize}

These results call into question theoretical methods that impose locality so as to
scale linearly with system size (LCCSD(T), LMP2), or that approximate
the dispersion energy using a pair-wise $-C_6 R^{-6}$ interaction, as is done
with dispersion-corrected DFT methods and empirical potentials. 
In the former, these effects can be included by extending the region of
locality, though at the cost of losing linearity in scaling, 
but the latter methods should not be applied to systems such as these.
Even DFT functionals with a non-local dispersion correction, such as the
van der Waals functional of Dion et al.\cite{DionRSLL04} and the more recent
functional of Vydrov and Van Voorhis \cite{VydrovVanV09a} are unlikely to contain
the correct physics because of an implicit assumption of locality in the
polarizability. These functionals will include many-body non-additive effects
between non-overlapping systems, but not the non-additive effects {\em within}
each system such as those described here.
On the other hand, methods based on the random phase approximation and
quantum Monte Carlo should be able to describe the non-additive effects described
in this paper if finite-size effects are kept under control.

In systems containing carbon atoms, such as $\pi$-conjugated chains, the contributions
from the core electrons will at least partially mask those from the more mobile
$\pi$ electrons, so it is possible that the changes in power-law of the
dispersion interaction will not be as dramatic as those we see in the \HH{n}
chains. But we should nevertheless expect a high degree of non-additivity arising
from the charge-flow terms. One indicator of this non-additivity is the
non-linear dependence of the molecular polarizability on system size. This has
been already demonstrated above and is further supported by the experimental work of 
Compagnon \etal \cite{CompagnonABDLR01} on the polarizabilities of the fullerenes
C70 and C60 which are found to be in the ratio 1.33 rather than the ratio
$70/60=1.17$ that would be expected if the systems were additive.
Therefore, if reliable $C_6$-models are to be constructed for extended $\pi$-conjugated
systems, these models will need to absorb the effects of the
non-additivity of the charge-flow polarizabilities. That is, they should be
tailored to suit the electronic structure of the system rather than be {\em
transferred} from calculations on smaller systems.
The Williams--Stone--Misquitta (WSM) procedure
\cite{MisquittaS08a,MisquittaSP08,MisquittaS08b} is one such method that is
capable of constructing effective local polarizability and dispersion models
that account for the electronic structure of the system. In fact, the
non-local models presented in this paper are derived in the first step of the
WSM procedure. We are currently investigating the behaviour of WSM dispersion
models for a variety of carbon systems.

Finally, strongly delocalized systems will also have important 
contributions to the dispersion energy from terms of third order in the
interaction operator, that is, from the hyperpolarizabilities.
We have not considered such terms in this paper. Furthermore, in ensembles of
such systems there will be strong non-additive effects {\em between} molecules.
We are currently investigating the nature and importance of this type of non-additivity.

\section{Acknowledgements}
We thank University College London for computational resources. AJM
and JS thank EPSRC for funding.

% Now include the BIBLIOGRAPHY.bib file that contains all references and uses
% the abbreviations from the above file.
\setlength\bibsep{2pt}
%\bibliography{BIBLIOGRAPHY}

\end{document}